\documentclass[reprint,aps,showpacs,pra,nofootinbib,superscriptaddress]{revtex4-2}

\usepackage{graphicx,newlfont}
\usepackage[dvipsnames]{xcolor}
\usepackage{physics,hyperref}
\usepackage{amssymb,amsmath,amsthm,mathtools}
\usepackage{bbm,braket,mathtools,mathrsfs}
\usepackage{bbold}
\usepackage{dblfloatfix} 
\usepackage[varg]{txfonts} 
\usepackage{lipsum}

\newcommand{\mf}{\mathfrak}
\newcommand{\mc}{\mathcal}
\newcommand{\ms}{\mathscr}
\newcommand{\mbb}{\mathbbm}
\newcommand{\eq}[1]{\begin{align}#1\end{align}}

\begin{document}
\title{Quantum violations of joint reality}
\author{R. A. Caetano}
\author{R. M. Angelo}
\affiliation{Department of Physics, Federal University of Paran\'a, P.O. Box 19044, 81531-980, Curitiba, Paran\'a, Brazil}

\begin{abstract} 
With basis on (i) the physical principle of local causality and (ii) a certain notion of elements of reality, Einstein, Podolsky, and Rosen (EPR) put forward an argument showing that physical instances may exist in which two non-commuting observables can be joint elements of the physical reality. Here, we introduce a new criterion of joint reality. We demonstrate that, according to this criterion, quantum mechanics generally prevents non-commuting observables from having joint elements of reality.
joint elements of reality. In addition, we introduce a measure to quantify the extent to which the criterion is violated and explore the implications of such a measure in connection with incompatibility and correlations. Our findings suggest new manners of interpreting quantum phenomena.
\end{abstract}

\maketitle

\section{Introduction}

The conflict between Quantum Mechanics (QM) and our everyday reality began to take shape in the early days of QM. The quantum principle of superposition, in which a linear combination of two legal states is also a legal state, is the primordial example of such conflict. Classical objects are localized, while quantum objects can be found in a superposition of positions. This contradiction is taken to the extreme in Schrödinger's cat thought experiment, where a cat is both alive and dead before one opens the box~\cite{schrodinger}. Einstein, along with many others, believed that QM was not a fundamental theory but merely a step toward a deeper theory of nature. The formalization of this idea was proposed in 1935 by Einstein, Podolsky, and Rosen (EPR)~\cite{EPR}. 

In their work, the authors introduced the following definition of \textit{element of reality}: ``If, without in any way disturbing a system, we can predict with certainty the value of a physical quantity, then there exists an element of physical reality corresponding to this physical quantity.'' Along with the premise that, in a \textit{complete theory}, ``every element of the physical reality must have a counterpart'' and the fact that QM does not predict definite values for noncommuting observables, EPR arrived at the dilemma: ``(1) the quantum mechanical description of reality given by the wave function is not complete, or (2) when the operators corresponding to two physical quantities do not commute, the two quantities cannot have simultaneous reality.'' By assuming local causality and arguing that there are quantum states for which noncommuting observables can be jointly real, EPR proved (2) wrong and claimed the incompleteness of quantum theory. To better appreciate EPR's point, let us consider, as a handy example, the singlet state $\text{\small $\sqrt{2}$}\ket{\psi_-}=\ket{0,1}-\ket{1,0}=\ket{+,-}-\ket{-,+}$ of two spin-1/2 particles, where $\{\ket{0},\ket{1}\}$ and $\{\ket{+},\ket{-}\}$ denote the eigenbasis of the operators $\sigma_z$ and $\sigma_x$, respectively. When measuring the particle spin in her possession either in the $z$ direction or in the $x$ direction, Alice can eventually project the state to either $\ket{0,1}$ or $\ket{+,-}$, respectively, thereby gaining full predictability about the spin value of the particle in Bob's distant laboratory without disturbing the remote particle. According to EPR's criterion, in this run, either $\sigma_z$ or $\sigma_x$ would be an element of reality in Bob's laboratory. Since Alice's choice cannot---by local causality---influence the reality at a distance, EPR would conclude in this scenario that both $\sigma_z$ and $\sigma_x$ must be elements of reality, even though their values are not concomitantly predicted by QM, which is then termed incomplete.

Already in 1935, Bohr addressed EPR's criticisms of QM by invoking the complementarity principle, where quantities associated with incompatible observables cannot be elements of reality in the same experimental arrangement~\cite{Bohr1935}. An important argument against the existence of local hidden variables was provided by Bell~\cite{bell1,bell2}, who demonstrated that a generic theory based on such variables would be inconsistent with the results predicted by QM. Although the ultimate meaning of Bell's hypothesis---especially with respect to some possible connection with realism---still inspires debate (see Ref.~\cite{PSilva2024} and references therein), very sophisticated loophole-free experiments \cite{hensen,giustina,shalm,hensen2,rauch,li} have demonstrated the inadequacy of models based on local hidden variables.

The emergence of physical reality has been addressed from different perspectives, ranging from models involving decoherence \cite{joos,zurek1,Schlosshauer} and quantum Darwinism \cite{zurek2} to approaches with more than one observer, akin to Wigner's friend arguments \cite{brukner,Frauchiger}. Even the very interpretation of the quantum state has come under scrutiny. On one hand, several works advocate in favor of the so-called $\psi$-ontic models, in which the wave function has a realistic nature \cite{pusey,lewis,colbeck,hardy,patra,aaronson,leifer,barret,branciard}. On the other hand, some authors defend the $\psi$-epistemic point of view, where the wave function represents knowledge of some underlying reality \cite{spekkens,Harrigan}. Taking a different approach, Bilobran and Angelo (BA) discuss realism from the perspective of the definite values of physical quantities~\cite{Bilobran2015}. The idea behind BA's criterion of reality is that a measurement of an observable establishes an element of reality for the measured observable, even if the result is not revealed to the observer. Thus, if a state is not altered by an unrevealed measurement, such a preparation already possessed an element of reality before the measurement. In the same work, BA introduce a quantifier---the so-called {\it irreality} (the complement of reality)---to measure how far a given preparation is  from a situation of full realism for a given observable. Irreality furnishes a gradation for quantum irrealism (violations of BA's realism) and allows one to diagnose how reality emerges dynamically upon interactions between quantum systems. This framework has been explored in subsequent works, yielding developments such as a novel notion of nonlocality~\cite{gomes1, orthey2019, fucci}, implications for foundational aspects of quantum theory~\cite{dieguez, engelbert,paiva2023} (including experimental tests~\cite{mancino, dieguez2}), and its role as a quantum resource~\cite{costa}. Recently, the concept of irreality has received a very formal treatment, in terms of axiomatization of its principles~\cite{Orthey2022}.

However versatile the BA approach may be, it remains elusive regarding a very fundamental aspect claimed in EPR's paper, namely, that joint elements of reality can manifest for noncommuting observables in scenarios involving strongly correlated systems. Although this conclusion sounds natural within the scope of classical physics, here we claim, based on a certain criterion of \textit{joint reality}, that QM generally implies the opposite. Recently, grounded in the hypothesis of operational completeness, the concept of joint reality and its (experimental) violation have been approached through device-independent no-go tests using linear and nonlinear inequalities for dichotomic observables~(see Ref.~\cite{Zhang2024} and references therein for other approaches to joint reality). Here, we take a fundamentally different route to joint reality, using the hypothesis of irrelevance of unrevealed measurements and introducing a quantifier of joint reality violations that applies not only to dichotomic observables. We extend the BA's criterion of reality to instances involving two observables, whether they are commuting or noncommuting. Most importantly, this is done without resorting in any way to Bell inequality violations, whose ultimate meaning does not clearly separate the notions of reality and locality. Then, using the von Neumann entropy, we introduce a quantifier of violations of this criterion and proceed with a thorough analysis of its physical implications. All this content is thoroughly presented in Sec.~\ref{sec:SI}. In Sec.~\ref{sec:examples}, we apply our framework to specific physical systems and show that the violations of joint reality are linked with the state information (purity), aspects of incompatibility, and correlations. Section~\ref{sec:conclusions} closes this article with our concluding remarks.

\section{Assessment of joint reality}
\label{sec:SI}

\subsection{Criterion and measure}

We start with the key idea introduced in Ref.~\cite{Bilobran2015}, namely,  that probability distributions remain unaltered by unrevealed measurements in realistic theories. To appreciate this point, let us consider the classical statistical mechanics of a particle moving in one dimension. The description of this system is given, in phase space, by the probability density $\wp_t(q,p)$, solution of the Liouville equations $\partial_t\wp_t=\{\mathrm{H},\wp_t\}$, with $\mathrm{H}$ the Hamiltonian function. A measurement of the generalized coordinate $q$ performed at time $t$, yielding outcome $\bar{q}$, results in the transition
\eq{\wp_t(q,p)\to \tilde{\wp}_t(q,p|\bar{q})\equiv\frac{\delta(q-\bar{q})\,\wp_t(\bar{q},p)}{\int\!\int dq\,dp\,\delta(q-\bar{q})\,\wp_t(\bar{q},p)}.
}
It's evident that the resulting description, with the revealed outcome $\bar{q}$, differs from the original one. Now, suppose the information about the outcome $\bar{q}$ is lost for some reason or is not revealed to the observer, who knows that a measurement has been performed. In this case, the best description available will be an average over all possible outcomes $\bar{q}$ weighted with the probability of its occurrence, $\mf{p}_t(\bar{q})=\int\! dp\,\wp_t(\bar{q},p)$. From this process of unrevealed measurement, it follows that
\eq{\label{unrev_measurement} \int\! d\bar{q}\, \tilde{\wp}_t(q,p|\bar{q})\,\mf{p}_t(\bar{q}) =\wp_t(q,p).
}
The re-obtainment of the original distribution is an expected result for a realistic theory like classical statistical mechanics. Indeed, in such a theory, the role of a measurement is merely to unveil a pre-established element of reality (such as $\bar{q}$, in this example); that is to say, the measurement does not materialize reality. Thus, when the outcome is kept secret, the measurement turns out to be innocuous, and nothing changes in the epistemic description. A second important aspect of a realistic theory, as noted long ago by Fine~\cite{Fine1982}, is the existence of joint probability distributions such as $\wp_t(q,p)$. It is evident from the arguments given above, that the same result would be obtained for unrevealed measurements of $p$, or even from sequential measurements of $q$ and $p$ in any order. 

We are now ready to introduce our criterion of joint reality. Consider a bipartite scenario where $\rho$ acts on a joint Hilbert space $\mc{H}\equiv\mc{H_A\otimes H_B}$ of dimension $d=d_\mc{A}d_\mc{B}$. Let $X=\sum_ix_iX_i$ and $Y=\sum_jy_jY_j$ be the spectral decompositions of nondegenerate discrete spectrum observables acting on $\mc{H}$, where $x_i$ and $y_j$ are eigenvalues, and $X_i$ and $Y_j$ are projection operators satisfying $X_iX_k=\delta_{ik}X_i$ and $Y_jY_k=\delta_{jk}Y_j$, respectively. An unrevealed measurement is here represented by a completely positive trace-preserving (CPTP) unital map, which is defined as
\eq{\label{Phi_X} \Phi_X(\rho)\coloneqq \sum_i X_i\,\rho\,X_i.
}
Using the forms $X_i=\ket{x_i}\!\bra{x_i}$ and $Y_j=\ket{y_j}\!\bra{y_j}$, one can show that $\Tr\big[Y_j\Phi_X(\rho)\big]=\sum_i\Tr\big[Y_jX_i\big]\Tr[X_i\rho]$, which can be interpreted as $p_{\Phi_X(\rho)}(y_j)=\sum_ip(y_j|x_i)p_{\rho}(x_i)$, in a faithful representation of the unrevealed measurement protocol leading to Eq.~\eqref{unrev_measurement}. In Ref.~\cite{Bilobran2015}, BA took the relation $\Phi_X(\rho)=\rho$ as criterion for realism (even when $X$ acts solely on $\mc{H_A}$), specifically meaning that $X$ is an element of reality for the preparation $\rho$. In this case, $\rho$ is referred to as an $X$-reality state. Then, BA introduced the {\it irreality} 
\eq{\label{mfI} \mf{I}_X(\rho)\coloneqq S\big(\Phi_X(\rho)\big)-S(\rho)=S\big(\rho||\Phi_X(\rho)\big),}
a measure of how much $X$ is unreal (uncertain) for the preparation $\rho$. Here, $S(\rho)=-\Tr(\rho\log\rho)$ denotes the von Neumann entropy of $\rho$, while $S(\rho||\sigma)=\Tr[\rho(\log{\rho}-\log{\sigma})]$ represents the relative entropy of the quantum states $\rho$ and $\sigma$.

In direct analogy with BA's approach, here we propose to adopt
\eq{\label{simult_reality} \Phi_{XY}(\rho)=\Phi_{YX}(\rho)=\rho
}
as a criterion of joint reality regarding the triple $\{X,Y,\rho\}$, where $\Phi_{RS}(\rho)\equiv \Phi_R\circ\Phi_S(\rho)$ denotes an ordered sequential  composition of unrevealed measurement maps for observables $R$ and $S$. Whenever condition \eqref{simult_reality} is fulfilled, the observables $X$ and $Y$ are said to be joint elements of reality for the state $\rho$. In this case, $\rho$ is referred to as an $\{X,Y\}$-reality state. In addition, because $\Phi_R^2=\Phi_R$, when Eq.~\eqref{simult_reality} holds, one has $\Phi_X(\rho)=\Phi_X^2\circ\Phi_Y(\rho)=\Phi_{XY}(\rho)=\rho$ and, similarly, $\Phi_Y(\rho)=\rho$, meaning that $X$ and $Y$ are also elements of reality individually, in BA's sense.

The second equality in criterion \eqref{simult_reality} implements the idea that unrevealed measurements are pointless when the measured observables are joint elements of reality. The first equality allows us to envisage joint probabilities for $X$ and $Y$. To illustrate this, let us consider the observable $Z=\sum_kz_kZ_k$, where $Z_k$ are projection operators. The probability of obtaining $z_k$ in a $Z$ measurement, given the states $\Phi_{XY}(\rho)$ or $\Phi_{YX}(\rho)$, is
\eq{\label{pzk1}\Tr[Z_k\Phi_{XY}(\rho)]=\sum_ip(z_k|x_i)\sum_jp(x_i|y_j)p_\rho(y_j)
}
or
\eq{\label{pzk2}\Tr[Z_k\Phi_{YX}(\rho)]=\sum_jp(z_k|y_j)\sum_ip(y_j|x_i)p_\rho(x_i),
}
respectively. In these relations we have used the interpretations $p(z_k|x_i)=\Tr[Z_kX_i]$, $p_\rho(x_i)=\Tr[X_i\rho]$, $p_\rho(y_j)=\Tr\big[Y_j\rho\big]$, and  $p(x_i|y_j)=p(y_j|x_i)=\Tr\big[Y_jX_i\big]$. Now, assumption \eqref{simult_reality} imposes that Eq.~\eqref{pzk1} and \eqref{pzk2} are one and the same. It follows that the second summations in each one of those equations have the connotation of marginal probability distributions arising from a joint probability given by 
\eq{p(x_i|y_j)\,p_\rho(y_j)=p(y_j|x_i)\,p_\rho(x_i)\equiv p(x_i,y_j).
}
Additionally, note that the validity of Bayes' rule (independence on the conditionalization order) is implied, which is expected to be the case when $[X,Y]=0$.

Having demonstrated the reasonableness of the criterion \eqref{simult_reality}, we now propose a quantifier to measure the degree of violation of this criterion. Although many forms are admissible in principle, we opt for a simple one, inspired by BA's work. We define the {\it joint irreality} (JI) as
\eq{\label{IXY}
\mf{I}_{\{X,Y\}}(\rho)\coloneqq\frac{S\big(\Phi_{XY}(\rho)\big)+S\big(\Phi_{YX}(\rho)\big)}{2}-S(\rho).}
Since the von Neumann entropy of $\rho$, $S(\rho)\coloneqq-\Tr[\rho \log{\rho}]$, is nondecreasing under CPTP maps (i.e., $S\big(\Phi_R(\rho)\big)\geqslant S(\rho)$), the above measure is guaranteed to be nonnegative. By construction, JI informs us that $X$ and $Y$ will be joint elements of reality if and only if criterion \eqref{simult_reality} is satisfied. To prove this assertion, we first add and subtract terms such as $S\big(\Phi_X(\rho)\big)$ and  $S\big(\Phi_Y(\rho)\big)$ to rewrite JI in the form
\eq{\label{mfIXY=sum}\mf{I}_{\{X,Y\}}(\rho)=\frac{\mf{I}_X(\rho)+\mf{I}_Y(\rho)+\mf{I}_X\big(\Phi_Y(\rho)\big)+\mf{I}_Y\big(\Phi_X(\rho)\big)}{2}.
}
As in Eq.~\eqref{mfI}, each irreality can be written in terms of the relative entropy, which is never negative and vanishes if and only if its arguments are equal. Then, having $\mf{I}_{\{X,Y\}}(\rho)=0$ requires that $\rho=\Phi_X(\rho)=\Phi_Y(\rho)=\Phi_{XY}(\rho)=\Phi_{YX}(\rho)$, which precisely corresponds to criterion \eqref{simult_reality}, thus completing the proof. 

\subsection{Implications}

An important consequence of the concept of JI is that, in general, far-apart systems do not always have joint elements of reality. Indeed, Eq.~\eqref{mfIXY=sum} reveals that JI encompasses more than just the sum of the irrealities, thus highlighting the fact that the concept here introduced is a nontrivial extension of BA's approach. The sum is verified only under very specific conditions, as we discuss now. Consider local observables, $X=A\otimes\mbb{1}$ and $Y=\mbb{1}\otimes B$, such that $[X,Y]=0$. In this case, JI reduces to 
\eq{\label{mfID} \mf{I}_{\{A\otimes \mbb{1},\mbb{1}\otimes B\}}(\rho)=\mf{I}_A(\rho_\mc{A})+\mf{I}_B(\rho_\mc{B})+D_{AB}(\rho),
}
where $D_{AB}(\rho)\coloneqq \ms{I}(\rho)-\ms{I}\big(\Phi_{AB}(\rho)\big)$ is the symmetric quantum discord of $\rho$ associated with the measurements $A$ and $B$~\cite{Rulli2011}, $\ms{I}(\rho)=S(\rho||\rho_\mc{A}\otimes\rho_\mc{B})$ is the mutual information of $\rho$, $\rho_\mc{A(B)}=\Tr_\mc{B(A)}[\rho]$ is the reduced state of part $\mc{A(B)}$, and $\Phi_{A B}(\rho)\equiv\Phi_{A\otimes \mbb{1}}\circ\Phi_{\mbb{1}\otimes B}(\rho)=\sum_{i,j}(A_i\otimes B_j)\rho (A_i\otimes B_j)$ is a bilocal map. More on the derivation and interpretation of Eq.~\eqref{mfID} can be found in Ref.~\cite{Savi2021}. What is noteworthy in Eq.~\eqref{mfID} is that quantum correlations captured by $D_{AB}(\rho)$ (of which entanglement is a particular case) prevent JI from being expressed solely in terms of the local irrealities $\mf{I}_A(\rho_\mc{A})$ and $\mf{I}_B(\rho_\mc{B})$. As an example, consider the singlet state $\psi_-=\ket{\psi_-}\!\bra{\psi_-}$, with $\ket{\psi_-}=(\ket{01}-\ket{10})/\text{\small $\sqrt{2}$}$, and set $A=B=\sigma_z$. Because $\rho_\mc{A,B}$ are maximally mixed states, it follows that $\mf{I}_{\sigma_z}(\rho_\mc{A,B})=0$. From $\Phi_{\sigma_z\sigma_z}(\psi_-)=\frac{1}{2}(\ket{01}\!\bra{01}+\ket{10}\!\bra{10})$, one obtains $D_{\sigma_z\sigma_z}=\log{2}$. Hence, $\mf{I}_{\{\sigma_z\otimes\mbb{1},\mbb{1}\otimes\sigma_z\}}=\log{2}$. This shows that even when $\mf{I}_A(\rho_\mc{A})=\mf{I}_B(\rho_\mc{B})=0$ (indicating realism of local observables), one may still have $\mf{I}_{\{A,B\}}(\rho)> 0$, implying joint irreality. In other words, for local observables, joint reality (the complement of JI) manifests itself as the sum of local realities only in the absence of quantum correlations. In particular, for instances involving no correlations whatsoever, one has
\eq{\mf{I}_{\{A\otimes \mbb{1},\mbb{1}\otimes B\}}(\rho_\mc{A}\otimes\rho_\mc{B})=\mf{I}_A(\rho_\mc{A})+\mf{I}_B(\rho_\mc{B}).
}

Let us now analyze a scenario where $[X,Y]\neq 0$. In particular, we consider local observables $X=A\otimes\mbb{1}$ and $Y=\bar{A}\otimes \mbb{1}$, where $\bar{A}$ is maximally incompatible with $A$, meaning that their eigenbases form maximally unbiased bases (MUB). In this case, we find $\Phi_{XY}(\rho)=\frac{\mbb{1}}{d_\mc{A}}\otimes \rho_\mc{B}$ and, hence,
\eq{ \mf{I}_{\{A\otimes\mbb{1},\bar{A}\otimes\mbb{1}\}}(\rho) = I(\rho_\mc{A})+\ms{I}(\rho),
} 
with $I(\rho_\mc{A})=\log{d}_\mc{A}-S(\rho_\mc{A})$ being the information encoded in the state $\rho_\mc{A}$. This result shows that both purity and correlations (whether classical or quantum) influence the JI.

Another interesting implication of definition~\eqref{IXY} is that $\mf{I}_{\{X,Y\}}(\mbb{1}/d)=0$ for all pairs $\{X,Y\}$. This assigns to the maximally mixed state the status of most classical state, the one for which absolutely all observables are elements of reality according to criterion \eqref{simult_reality}. On the other hand, when $Y\equiv\bar{X}$, denoting an operator maximally incompatible with $X$, one shows that $\Phi_{X\bar{X}}(\rho)=\Phi_{\bar{X}X}(\rho)=\mbb{1}/d$ for all $\rho$. It follows from definition~\eqref{IXY} that $\mf{I}_{\{X,\bar{X}\}}(\rho)=\log{d}-S(\rho)=I(\rho)$, which maximizes JI, as demonstrated bellow. Notice, additionally, that in this case JI is equivalent to the total information encoded in $\rho$, which is also a measure of purity. Finally, it is noteworthy to examine how our approach leads to a conclusion that is contrary to EPR’s. As discussed in the Introduction, for the singlet state $\psi_-$, the EPR approach predicts that $\sigma_z$ and $\sigma_x$ are joint elements of reality in Bob's laboratory. Setting $X=\mbb{1}\otimes\sigma_z$ and $Y=\bar{X}=\mbb{1}\otimes\sigma_x$, we can use the above results to arrive at $\mf{I}_{\{\mbb{1}\otimes\sigma_z,\mbb{1}\otimes\sigma_x\}}(\psi_-)=I(\psi_-)=\log{4}$, implying maximal violation of joint reality.

\subsection{Bounds}

Regardless of the choices of $X$ and $Y$, the nonnegativity of the irreality allows us to use Eq.~\eqref{mfIXY=sum} to write $2\mf{I}_{\{X,Y\}}(\rho)\geqslant \mf{I}_X(\rho)+\mf{I}_Y(\rho).$ This shows that JI can never be smaller than the average irreality. Using the entropic uncertainty relation $S(\Phi_X(\rho))+S(\Phi_Y(\rho)) \geqslant -2\log{\mf{c}}$ \cite{maassen} , with $\mf{c}=\max_{i,j}|\braket{x_i|y_j}|$, we finally arrive at $\mf{I}_{\{X,Y\}}(\rho)\geqslant -\log{\mf{c}}-S(\rho)$. An upper bound for JI can be found from Eq.~\eqref{IXY} by noticing that the von Neumann entropy is never greater than $\log{d}$. It follows that $\mf{I}_{\{X,Y\}}(\rho)\leqslant \log{d}-S(\rho)=I(\rho)$. Putting all this together yields
\eq{\label{bounds} I(\rho)-\log{(\mf{c} d)}\leqslant \mf{I}_{\{X,Y\}}(\rho)\leqslant I(\rho).
}
Clearly, the lower bound is not tight for maximally mixed state, in which case one has $\mf{I}_{\{X,Y\}}(\mbb{1}/d)=I(\mbb{1}/d)=0$. For pure states, $\psi=\ket{\psi}\!\bra{\psi}$, and maximally incompatible observables ($\mf{c}=1/\sqrt{d}$), one finds $\frac{1}{2}\log{d}\leqslant \mf{I}_{\{X,Y\}}(\psi)\leqslant \log{d}$.

\section{Case studies}
\label{sec:examples}

\subsection{joint irrealty for a qubit}

\begin{figure}[htb]
\centerline{\includegraphics[scale=0.48]{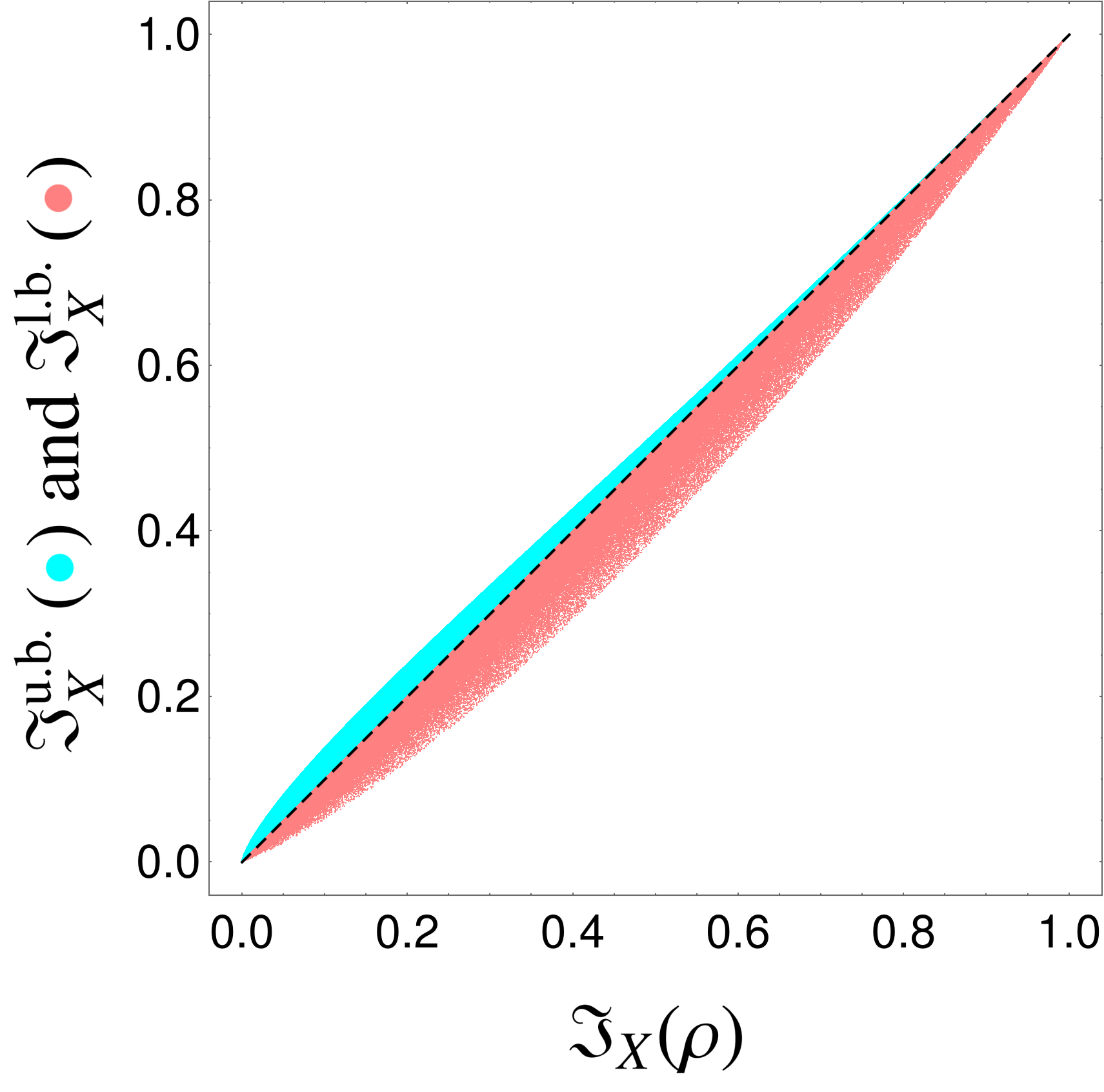}}
\caption{Parametric plot of the upper bound $\mf{I}_X^\text{u.b.}$ (cyan dots) and the lower bound $\mf{I}_X^\text{l.b.}$ (pink dots) [see Eq.~\eqref{qubit-bounds}] against the irreality $\mf{I}_X(\rho)$ for $2\times 10^6$ randomly generated configurations $(r,\lambda)$ within the range of $[0,1]$ (half of the configurations for each color), where $r$ is the norm of the vector $\vec{r}=r\hat{r}$ that defines $\rho$ on the Bloch sphere and $\lambda=|\hat{x}\cdot\hat{r}|$. Note that the deviation of $\mf{I}_X^\text{u.b.}$ from $\mf{I}_X(\rho)$ is fairly small. Since logarithms have been computed in base 2, the bounds and the irreality are bounded to 1 and given in bits. The dashed black line corresponds to the scenario in which $\mf{I}_X^\text{u.b.}=\mf{I}_X^\text{l.b.}=\mf{I}_X(\rho)$.}
\label{fig1}
\end{figure}

Consider the operators $\rho=\frac{1}{2}(\mbb{1}+\vec{r}\cdot\vec{\sigma})$, $X=\hat{x}\cdot\vec{\sigma}$, and $Y=\hat{y}\cdot\vec{\sigma}$ acting on a Hilbert space $\mc{H}$ of dimension $d=2$, where $\vec{r}$, $\hat{x}$, and $\hat{y}$ are vectors in the Bloch sphere (in $\mbb{R}^3$), with the latter two having unit norm, and $\vec{\sigma}$ is the vector whose components are the Pauli matrices. Via direct calculations one shows that $\Phi_X(\rho)=\frac{1}{2}[\mbb{1}+(\hat{x}\cdot\vec{r}\,\hat{x})\cdot\vec{\sigma}]$. Using the eigenvalues $(1\pm r)/2$ of $\rho$, one can express its von Neumann entropy as $S(\rho)=h[(1+r)/2]$, where $h(u)\coloneqq -u\log{u}-(1-u)\log(1-u)$ stands for the binary entropy of the probability distribution $\{u,1-u\}$, and $r=||\vec{r}||=(\vec{r}\cdot\vec{r})^{1/2}$. Hence, $S\big(\Phi_X(\rho)\big)=h[(1+r\lambda)/2]$, with $\lambda=|\hat{x}\cdot\hat{r}|$. After some numerical experimentation with the irreality $\mf{I}_X(\rho)=h\left(\tfrac{1+\lambda r}{2}\right)-h\left(\tfrac{1+r}{2}\right)$, we have found that $\mf{I}_X^\text{l.b.}\leqslant \mf{I}_X(\rho)\leqslant  \mf{I}_X^\text{u.b.}$ with the following bounds:
\begin{subequations}\label{qubit-bounds}
\eq{\mf{I}_X^\text{l.b.}&= I(\rho)\,(1-\lambda^2),\\
\mf{I}_X^\text{u.b.}&= I(\rho)\,(1-\lambda^2)^{3/4},}
\end{subequations}
where $I(\rho)=\log{2}-h[(1+r)/2]$. This result is numerically demonstrated by the parametric plot shown in Fig.~\ref{fig1} for $2\times 10^6$ randomly generated configurations $(r,\lambda)$ within the range of $[0,1]$. In particular, we see that $\mf{I}_X^\text{u.b.}(\rho)$ turns out to be a very good approximation of $\mf{I}_X(\rho)$. The mathematical structure of the bounds \eqref{qubit-bounds} raises the question whether we may validate the formula $\mf{I}_X(\rho)=I(\rho)\,(1-\lambda^2)^\mu$ for some real parameter. Via direct computation of $\mu$ through the inversion of this formula, we have succeeded to find a real value $\mu=\mu(X,\rho)\in (0.7, 1.0]$ for each pair $\{X,\rho\}$ of the $10^6$ analyzed.

Now, from well-known properties of the Pauli matrices, one computes $[X,\rho]=i(\hat{x}\times\vec{r})\cdot\vec{\sigma}$. Then, utilizing the Schatten 2-norm of a bounded operator $O$, $||O||_2=(\Tr\big[O^\dag O\big])^{1/2}$, one can estimate the degree of incompatibility between $X$ and $\rho$ through the formula $||[X,\rho]||_2=\sqrt{2}\,||\hat{x}\times\vec{r}||$. Because $|\hat{x}\cdot\hat{r}|^2+||\hat{x}\times\hat{r}||^2=1$, we can rewrite the proposed model in the form
\eq{\label{mfI-Inc}\mf{I}_X(\rho)=I(\rho)\,||\hat{x}\times\hat{r}||^{2\mu}=I(\rho)\left( \frac{||[X,\rho]||_2}{r\sqrt{2}}\right)^{2\mu},}
with $\mu=\mu(X,\rho)$. Although derived specifically for a qubit, this is a remarkable result: basically, it shows that irreality arises from two conceptually distinct factors, namely, state purity (information) and incompatibility. In other words, the irreality of $X$ given $\rho$, per unit of information, can be seen as a measure of incompatibility\footnote{Well developed as it is, the literature regarding the notion of incompatibility extends to instances far beyond mere non-commutativity~\cite{Heinosaari2016,Guhne2023}.} between $X$ and $\rho$. To a certain extent, this could have been anticipated based on the fact that $\rho=\Phi_X(\rho)$ (realism) immediately implies $[X,\rho]=0$ (compatibility). With result \eqref{mfI-Inc}, Eq.~\eqref{mfIXY=sum} can be re-expressed as a sum of terms involving the Schatten 2-norm of the commutators $[X,\rho]$, $[Y,\rho]$, $[X,\Phi_Y(\rho)]$, and $[Y,\Phi_X(\rho)]$, which highlights the various levels\footnote{It is interesting to note that the terms $\mf{I}_X\big(\Phi_Y(\rho)\big)$ and $\mf{I}_Y\big(\Phi_X(\rho)\big)$, appearing in Eq.~\eqref{mfIXY=sum} and related to $[X,\Phi_Y(\rho)]$ and $[Y,\Phi_X(\rho)]$, respectively, have been assigned the interpretation of {\it context incompatibility}~\cite{Martins2020}, a resource in communication tasks. It has been shown that, in the proper limit, context incompatibility retrieves measurement incompatibility.} of incompatibility behind JI. Moreover, these terms can be simply written (using the underlying geometry of the Bloch sphere) as functions of $||\hat{x}\times\hat{r}||$, $||\hat{y}\times\hat{r}||$, $||\hat{x}\times(\hat{y}\cdot\hat{r}\hat{y})||$, and $||\hat{y}\times(\hat{x}\cdot\hat{r}\hat{x})||$, respectively.

Let us specialize the discussion to the case where $X$ and $Y$ are maximally incompatible, in which case we have already seen that $\mf{I}_{\{X,Y\}}(\rho)=I(\rho)$. From $[X,Y]=2i(\hat{x}\times\hat{y})\cdot\vec{\sigma}$ and $||[X,Y]||_2=\text{\small $\sqrt{8}$}||\hat{x}\times\hat{y}||$, we conclude that the maximum incompatibility occurs for ``orthogonal operators''. Plugging the model \eqref{mfI-Inc} into Eq.~\eqref{mfIXY=sum} allows us to write
\eq{\frac{2\mf{I}_{\{X,Y\}}(\rho)}{I(\rho)}=||\hat{x}\times\hat{r}||^{2\mu_1}+||\hat{y}\times\hat{r}||^{2\mu_2}+|\hat{x}\cdot\hat{r}|^{2\mu_3}+|\hat{y}\cdot\hat{r}|^{2\mu_4},\nonumber
}
which furnishes the expected expression only if $\mu_{1,2,3,4}=1$. On the other hand, in the scenario where $[X,Y]=0$, which means that $||\hat{x}\times\hat{y}||=0$, we find
\eq{\frac{2\mf{I}_{\{X,Y\}}(\rho)}{I(\rho)}&=||\hat{x}\times\hat{r}||^{2\mu_1}+||\hat{y}\times\hat{r}||^{2\mu_2}\nonumber \\
&=\left( \frac{||[X,\rho]||_2}{r\sqrt{2}}\right)^{2\mu_1}+\left( \frac{||[Y,\rho]||_2}{r\sqrt{2}}\right)^{2\mu_2}.\nonumber
}
In analogy with Eq.~\eqref{mfID}, one sees that JI manifests for the pair $\{X,Y\}$ only if none of these operators commute with $\rho$.

\subsection{joint irreality for a two-qubit system}

\begin{figure}[htb]
\centerline{\includegraphics[width=\columnwidth]{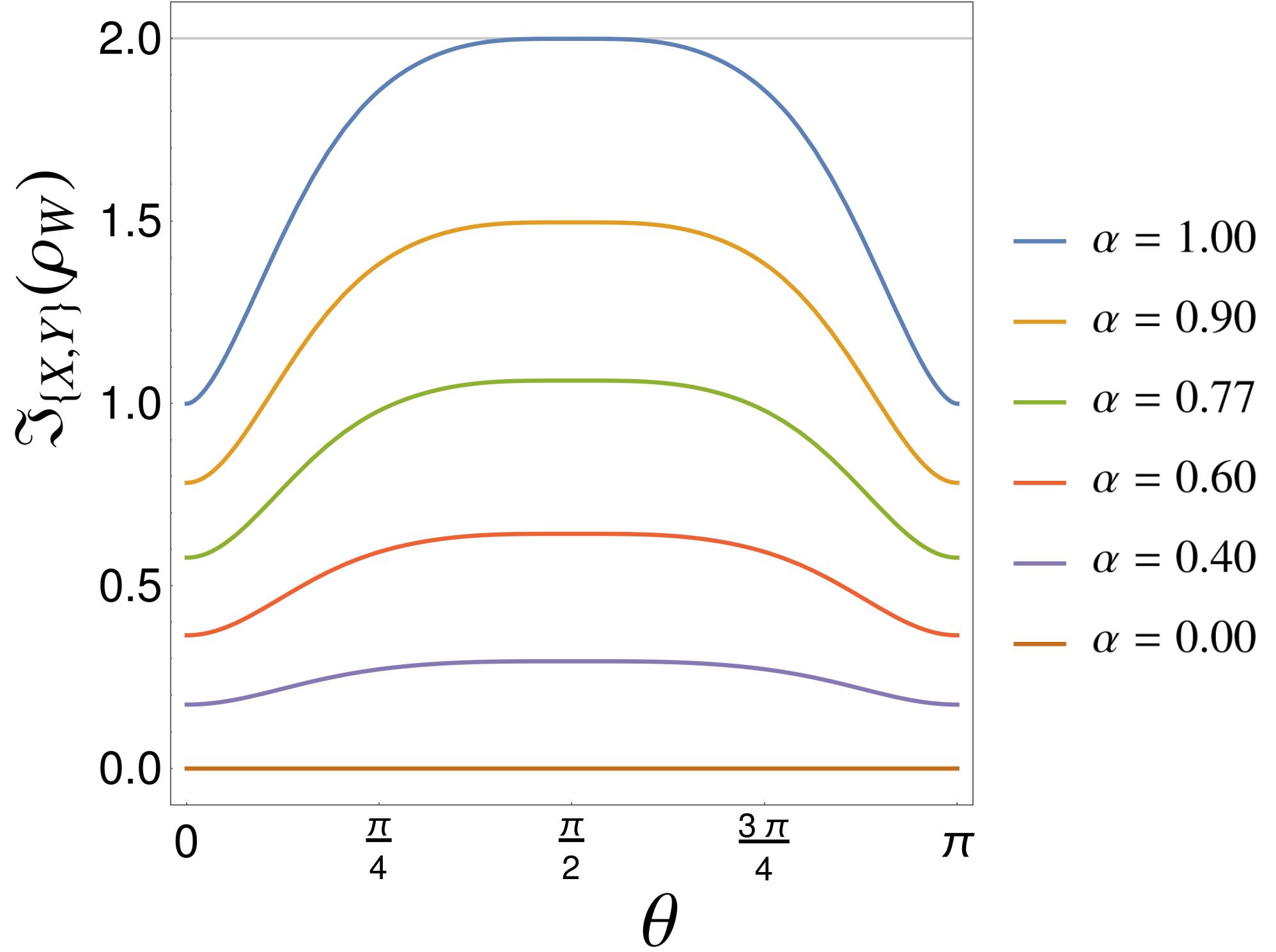}}
\caption{joint irreality $\mf{I}_{\{X,Y\}}(\rho_W)$ as a function of $\theta\in[0,\pi]$ for the Werner state $\rho_W$ with several values of the purity parameter $\alpha$ (the greater the value of this parameter, the higher the curve). Except for the totally mixed state ($\alpha=0$, bottom curve), where JI is always zero, for any other state with any degree of purity, irreality increases with $\theta$. Since logarithms have been computed in base 2, maximum irreality is bounded to $\log{d}=2$.}
\label{fig2}
\end{figure}

Up to now, we have been focused in situations where the two observables effectively acted either on the same Hilbert space or in completely different Hilbert spaces. Now we consider a two-qubit system and focus on the operators $X=\sigma_z\otimes \mbb{1}$ and $Y=(\sigma_x\cos{\theta}+\sigma_z\sin{\theta})\otimes\sigma_y$, with $\theta\in[0,\pi]$. That is, while $Y$ acts on the composite Hilbert space $\mc{H_A\otimes H_B}$, $X$ effectively acts only on $\mc{H_A}$. From $[X,Y]=(2i\cos{\theta})\sigma_{y}\otimes\sigma_y$, one obtains the Schatten estimator of incompatibility $||[X,Y]||_2=4|\cos{\theta}|$, which is clearly regulated only by $\theta$. The preparation under consideration will be the Werner state~\cite{Werner1989} $\rho_W=(1-\alpha)\frac{\mbb{1}}{4}+\alpha\psi_\pm$, where $\alpha\in[0,1]$, $\psi_\pm=\ket{\psi_\pm}\!\bra{\psi_\pm}$, and $\ket{\psi_\pm}=(\ket{0,1}\pm\ket{1,0})/\text{\small $\sqrt{2}$}$ (Bell states). Note that the real parameter $\alpha$ is a measure of state purity, as $I(\rho)$. After a tedious calculation, one obtains a function of $\alpha$ and $\theta$ for JI, which can be compactly expressed as
\eq{\mf{I}_{\{X,Y\}}(\rho_W)&=\tfrac{1}{2}\sum_{\epsilon=\pm 1}\left[3\,G\left(\nu_\epsilon\right)+G\left(\lambda_\epsilon\right)\right] \nonumber \\
&+\tfrac{1}{4} G(1+3\alpha)+\tfrac{3}{4}G(1-\alpha)-2\log{2},
}
where $G(u)\coloneqq -u \log{u}$, $\nu_\epsilon\equiv\{2+\epsilon \alpha[1+\cos(2\theta)]\}/8$, and $\lambda_\epsilon\equiv\{4+\epsilon \alpha[1+2\cos(2\theta)+\cos(4\theta)]\}/16$. In Fig.~\ref{fig2}, JI is plotted as a function of $\theta$ for several values of the purity parameter $\alpha$ (the greater the value of $\alpha$, the higher the curve in the graph).

\begin{figure}[htb]
\centerline{\includegraphics[scale=0.46]{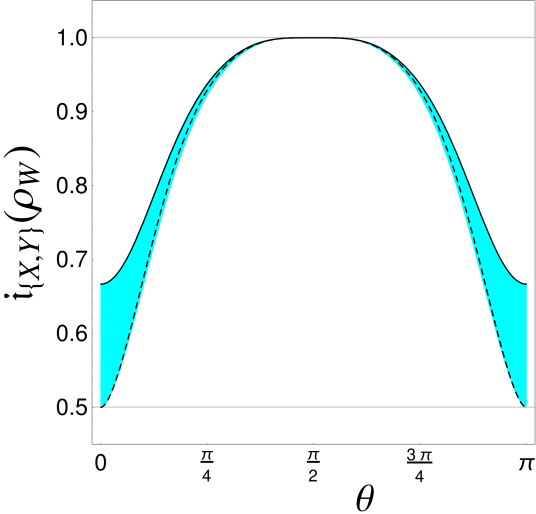}}
\caption{joint irreality per unit information, $\mf{i}_{\{X,Y\}}(\rho_W)=\mf{I}_{\{X,Y\}}(\rho_W)/I(\rho_W)$, as a function of $\theta\in[0,\pi]$ for the Werner state $\rho_W$ with $10^3$ randomly generated values of the purity parameter $\alpha$ (curves forming the smooth cyan surface). The solid black line refer to $\alpha\to 0$, whereas the dashed black line refers to $\alpha\to 1$. It is evident that the ``JI density'' never vanishes in the investigated domain: $\mf{i}_{\{X,Y\}}(\rho_W)\geqslant 1/2$.}
\label{fig3}
\end{figure}

Some aspects are readily understood in this result. First, for $\alpha=0$, the Werner state becomes a full mixture (classical state), for which all observables are expected to be elements of reality jointly. Second, in analogy with the results obtained in the previous section for the qubit [see Eq.~\eqref{mfI-Inc}], here we also have a dependence with the total information $I(\rho)$ (monotonically associated withe the purity parameter $\alpha$. As the curves in Fig.~\ref{fig1} suggest, JI goes to zero along with the information encoded in the state. The question then arises whether the JI per unit information, $\mf{i}_{\{X,Y\}}(\rho)\coloneqq \mf{I}_{\{X,Y\}}(\rho)/I(\rho)$, still vanishes when $\alpha\to 0$. The answer, for the case under inspection, is ``no'', as can be checked in Fig.~\ref{fig3} (see, in particular, the solid black line corresponding to $\alpha=0$). Also noteworthy is the fact that $\mf{i}_{\{X,Y\}}(\rho_W)=1$, for all $\alpha$, as $\theta=\pi/2$. The reason why this happens is that, at this point, $X$ and $Y$ become maximally incompatible, which implies the previously discussed result that $\mf{I}_{\{X,\bar{X}\}}(\rho)=I(\rho)$ for all $\rho$.

What is perhaps less intuitive is the fact that JI increases with $\theta$ in the domain $[0,\frac{\pi}{2}]$, while the commutativity of $X$ and $Y$ diminishes (with inverse behaviors occurring in the domain $[\frac{\pi}{2},\pi]$). The flaw in this rationale comes from associating JI with the joint measurability of the observables. Actually, the association of Eq.~\eqref{mfI-Inc} with \eqref{mfIXY=sum} shows that what really matters are the relations $||[X,\rho_W]||=||[Y,\rho_W]||=\sqrt{2} \alpha$ and
\eq{\label{sino} ||[Y,\Phi_X(\rho_W)]||=\alpha |\sin{\theta}|,
}
while $|| [X,\Phi_Y(\rho_W)]||=0$. It is clear that, although $[X,Y]=0$ for $\theta=\pi/2$, $Y$ is maximally incompatible with $\Phi_X(\rho_W)$. Via Eq.~\eqref{sino}, we infer that the irreality $\mf{I}_Y\big(\Phi_X(\rho_W)\big)$ monotonically relates, as does $\mf{I}_{\{X,Y\}}(\rho_W)$, with some power of $|\sin{\theta}|$ rather than with with $|\cos{\theta}|$.

\subsubsection{The role of correlations}

For the scenario under scrutiny, we also have the opportunity to study the role of correlations in JI. Relation \eqref{mfID} already anticipates a link with the well-known symmetric quantum discord $D_{AB}(\rho)$ of the local measurements $A$ and $B$. However, to the best of our knowledge, the current literature does not provide us with a well-established estimator of correlations for scenarios where $X$ and $Y$ are both acting on $\mathcal{H}_A \otimes \mathcal{H}_B$ (nonlocal observables). We then propose to consider the following measure:
\eq{\label{delta} \delta_{X,Y}(\rho)\coloneqq \big|\ms{I}(\rho)+\ms{I}\big(\Phi_{XY}(\rho_\mc{A}\otimes\rho_\mc{B})\big)-\ms{I}\big(\Phi_{XY}(\rho)\big)\big|,
}
where $\ms{I}$ is the mutual information and $\rho_\mc{A,B}$ are reduced states. This measure is a clear generalization of the $AB$ symmetric quantum discord, since $\delta_{A\otimes\mbb{1},\mbb{1}\otimes B}(\rho) = D_{AB}(\rho)$. Moreover, it is readily seen that $\delta_{X,Y}(\rho_{\mc{A}} \otimes \rho_{\mc{B}}) = 0$ for all $X$ and $Y$, showing that this measure is a faithful estimate of correlations. The modulus is introduced to ensure the nonnegativity of the measure. Since $X$ and $Y$ do not commute in general, it makes sense to consider the symmetrized form
\eq{\label{mcD}\mc{D}_{\{X,Y\}}(\rho)\coloneqq \frac{\delta_{X,Y}(\rho)+\delta_{Y,X}(\rho)}{2}.
}
Although there are currently no theoretical grounds to declare $\mc{D}_{\{X,Y\}}(\rho)$ as a definitive measure of correlations (whether classical or quantum) for nonlocal observables, we consider it a reasonable estimate for the present purposes.

\begin{figure}[htb]
\centerline{\includegraphics[scale=0.45]{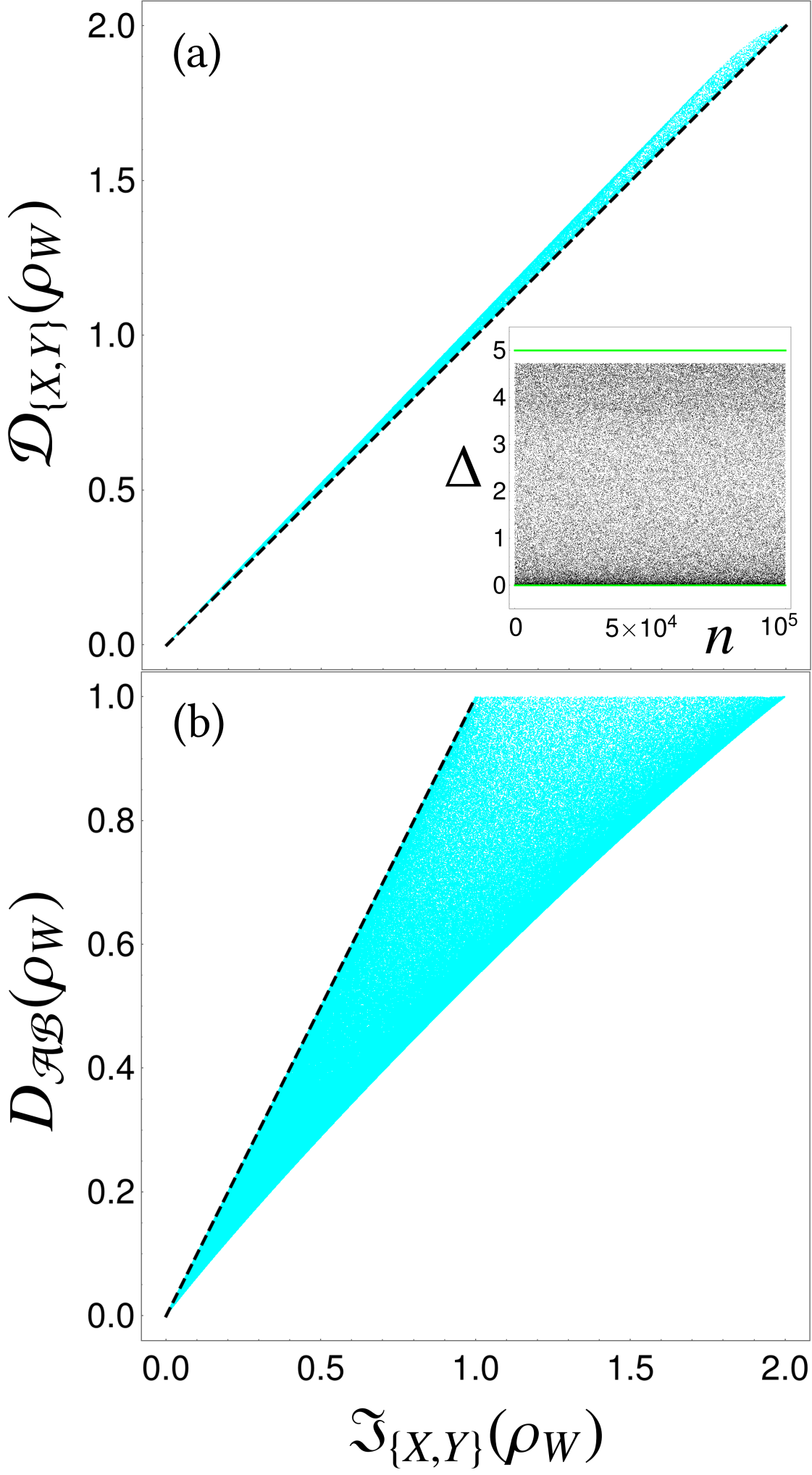}}
\caption{(a) Parametric plot of $\mc{D}_{{X,Y}}(\rho_W)$ against $\mf{I}_{{X,Y}}(\rho_W)$, for the Werner state $\rho_W$. Each one of the $10^5$ cyan dots corresponds to a randomly generated triple $\{X,Y,\rho_W\}$, parametrized by the pair $\{\alpha,\theta\}$ (see main text). The dashed black line refers to instances where $\mc{D}_{{X,Y}}(\rho_W)=\mf{I}_{{X,Y}}(\rho_W)$. Inset: Each black dot corresponds to the percentual deviance $\Delta\equiv 100\left[\mc{D}_{{X,Y}}(\rho_W) - \mf{I}_{{X,Y}}(\rho_W)\right]/\mf{I}_{{X,Y}}(\rho_W)$ as a function of the $n$-th triple $\{X,Y,\rho_W\}$, with $1\leqslant n\leqslant 10^5 \in \mbb{N}$. The green solid lines inform that $0 \leqslant \Delta < 5$. These results allow us to conclude that  $\mf{I}_{\{X,Y\}}(\rho_W)\lessapprox \mc{D}_{\{X,Y\}}(\rho_W)$. (b) Parametric plot of the average one-sided quantum discord $D_\mc{AB}(\rho_W)$ against $\mf{I}_{{X,Y}}(\rho_W)$, for the Werner state $\rho_W$. Each one of the $10^5$ cyan dots corresponds to a randomly generated triple $\{X,Y,\rho_W\}$, parametrized by the pair $\{\alpha,\theta\}$. The simulations showed that $\mf{I}_{{X,Y}}(\rho_W)\geqslant D_\mc{AB}(\rho_W)$, with equality (dashed black line) holding for $(\alpha,\theta)=(\forall,k\pi)$, with $k\in\{0,1\}$, and $(\alpha,\theta)=(0,\forall)$.}
\label{fig4}
\end{figure}

Direct calculations for the Werner state show that $\rho_\mc{A,B}^W=\frac{\mbb{1}}{2}$, so that $\ms{I}\big(\Phi_{XY}(\rho_\mc{A}^W\otimes\rho_\mc{B}^W)\big)=0$. Although analytical, the result for $\mc{D}_{\{X,Y\}}(\rho_W)$ is cumbersome and will be omitted. In Fig.~\ref{fig4}(a) we present a numerical comparison between $\mc{D}_{\{X,Y\}}(\rho^W)$ and $\mf{I}_{\{X,Y\}}(\rho_W)$ for $10^5$ random realizations of the triple $\{X,Y,\rho_W\}$ (through the parameters $\alpha$ and $\theta$). In the inset, we see the percentual deviance $\Delta\equiv 100\,[\mc{D}_{\{X,Y\}}(\rho_W)-\mf{I}_{\{X,Y\}}(\rho_W)]/\mf{I}_{\{X,Y\}}(\rho_W)$ for each run of the simulation.

To enhance the connection of JI with exclusively quantum correlations, we also computed the minimized one-sided quantum discord~\cite{Vedral2001,Zurek2002}, $D_\mc{A}(\rho)\coloneqq \min_A[\ms{I}(\rho)-\ms{I}\big(\Phi_A(\rho)\big)]$, a very well-established measure. In fact, here we use the average version $D_\mc{AB}(\rho)\coloneqq [D_\mc{A}(\rho)+D_\mc{B}(\rho)]/2$. For the Werner, known results~\cite{Costa2013,Costa2016} indicate that
\eq{\label{DmcAB} D_\mc{AB}(\rho_W)=\frac{2G(1+\alpha)-G(1-\alpha)-G(1+3\alpha)}{4}.
}
Naturally, the result does not depend on $\theta$. The comparison between the average one-sided quantum discord and JI is provided in Fig.~\ref{fig4}(b) for $10^5$ randomly generated pairs $\{\alpha,\theta\}$. All in all, the results shown in Fig.~\ref{fig4}
\eq{\label{SI-Disc}D_\mc{AB}(\rho_W)\leqslant \mf{I}_{\{X,Y\}}(\rho_W)\lessapprox \mc{D}_{\{X,Y\}}(\rho_W).
}
To a fairly good approximation, one may argue that the JI of $X$ and $Y$ for $\rho_W$ is induced by the correlations diagnosed by the measure~\eqref{mcD} and, in addition, is bounded from bellow by genuine quantum correlations.

\section{Concluding remarks}
\label{sec:conclusions}

Reconciling the theoretical framework of QM with our classical perception of nature has been a formidable problem since the inception of the theory. Quantum resources, such as coherence and entanglement, introduce fundamental aspects of indefiniteness that prevent us from envisioning physical quantities as elements of reality. In contrast to this perspective, EPR resorted to the principle of relativistic causality to conclude that noncommuting observables can jointly represent elements of reality for a given preparation \cite{EPR}. Traditional approaches attempting to refute EPR's conclusions often rely on violations of Bell inequalities, yet these results do not definitively indicate which concept---realism or locality---is to be rejected. Here, we conduct the discussion in terms of a well-motivated criterion of joint reality [Eq.~\eqref{simult_reality}], which has no conceptual dependence on locality, and then we demonstrate how QM violates this criterion. To quantify the degree of violation, we introduce the measure called JI [Eq.~\eqref{IXY}].

As immediate consequences of the introduced concepts, we have shown that (i) JI is a nontrivial generalization of the well-established concept of irreality~\cite{Bilobran2015}, and (ii) in the presence of correlations, local observables will not be elements of reality. We have also derived some bounds for JI and conducted two case studies. In the first one, for a qubit, we have demonstrated the link between irreality with information and incompatibility [Eq.~\eqref{mfI-Inc}]. Besides deepening the knowledge associated with irreality, this formula allows us to identify the various degrees of incompatibility encompassed in JI. Our second case study focused on a two-qubit system described by a triple $\{X,Y,\rho_W\}$, with $Y$ (resp. $X$) being a nonlocal (resp. local) observable and $\rho_W$ representing a Werner state. An important result in this framework is that JI goes to zero along with (and only because of) the state information, in a way such that JI per unit information never vanishes. This suggests that, typically, we should not expect two generic observables to be joint elements of reality. A second important contribution refers to the introduction of a measure of correlations involving nonlocal observables [Eq.~\eqref{delta}] and the observation that this measure is a very tight upper bound to JI. Moreover, we have shown that JI is lower bounded by the average one-sided minimized quantum discord, a genuine measure of quantum correlations. These results, expressed by the relations \eqref{SI-Disc}, constitute indirect evidence that JI keeps some dependence with the correlations encoded in the state.

In proposing a framework that is conceptually independent of the notion of locality, we hope that our findings may add evidence against local realism. Furthermore, the present work offers a relevant perspective concerning the understanding of the concept of realism, from the quantum substratum, and offers a new and relevant tool to interpret quantum phenomena.

\begin{acknowledgments}
R.M.A. acknowledges support of the National Institute for the Science and Technology of Quantum Information (INCT-IQ), Grant No. 465469/2014-0, and the National Council for Scientific and Technological Development (CNPq), Grant No. 305957/2023-6.
\end{acknowledgments}


\end{document}